\documentclass{article}

\begin{document}
\vspace*{-1.8cm}
\begin{flushright}
{\bf LAL 06-140}\\
\vspace*{0.1cm}
{September 2006}
\end{flushright}
\vspace*{1.5cm}

\begin{center}
{\LARGE\bf About the connection between vacuum birefringence\\
\vspace*{0,3cm}
 and the light-light scattering amplitude}
\end{center}
\vskip 1.0truecm

\begin{center}
{\bf\Large J. Ha\"{\i}ssinski, S. Dagoret, M. Urban and F. Zomer}\\
\vspace*{0,5cm}
{\bf\large Laboratoire de l'Acc\'el\'erateur Lin\'eaire,}\\
{\it IN2P3-CNRS et Universit\'e de Paris-Sud 11, BP 34, 
F-91898 Orsay Cedex}
\end{center}
\vspace*{0.5cm}

\begin{abstract}
\noindent
Birefringence phenomena stemming from \emph{vacuum} polarization are revisited in the framework of coherent scattering. Based on photon-photon scattering, our analysis brings out the direct connection between this process and vacuum birefringence. We show how this procedure can be extended to the Kerr and the Cotton-Mouton birefringences in vacuum, thus providing a unified treatment of various polarization schemes, including those involving static fields.
\end{abstract}

\vspace*{1cm}
\section {Introduction}

\indent
Non-linear optical effects stemming from the presence of virtual electron-positron pairs in vacuum are a fundamental feature of quantum field theory~\cite{Euler}. Among them, vacuum birefringence produced by a static electromagnetic field has been studied since the late fifties (for references of early works, see~\cite{First1}) and, later on, other non-linear effects have been considered, such as second-harmonic generation~\cite{KapDing}, generation of interacting modes in wave-guides~\cite{Brodin}, or self-interaction effects between photons~\cite{Brodin}.\\

This paper focuses on vacuum birefringence. Besides the Kerr birefringence due to an electric field and the Cotton-Mouton birefringence due to a magnetic one, magnetoelectric birefringences have been predicted by Rikken and Rizzo~\cite{Rikken1,Rizzo1} to take place in vacuum when crossed electric and magnetic fields are applied either perpendicularly or parallel to the direction of light propagation. Vacuum birefringence produced by an intense laser radiation field was previously investigated by Aleksandrov \emph{et al.}~\cite{Aleksandrov}.\\

Experiments -- either ongoing or in preparation -- aim at a direct detection of these very small effects and this goal is usually combined with an axion search~\cite{ongoing}. Prospects for an increase of the experimental sensitivity are also being investigated~\cite{Luiten, Heinze}.\\

Most authors base their calculations of optical non-linear effects in vacuum -- including birefringence  -- on the Heisenberg-Euler effective Lagrangian~\cite{Euler}

\begin{equation}
L = \frac{1}{2} (\rm{\mathbf{E}}^2 - \rm{\mathbf{B}}^2) + \frac{2}{45} \frac{\alpha^2 \hbar^3}{m_e^4 c^5} [(\rm{\mathbf{E}}^2 - \rm{\mathbf{B}}^2)^2 + 7(\rm{\mathbf{E}}\cdot \rm{\mathbf{B}})^2].
\end{equation}

With apologies to our younger readers, we  use Heaviside (rationalized) units as is commonly done in this field. Furthermore, we will set $\hbar=c=1$ below.\\

The connection between vacuum birefringence and light-light scattering is often mentioned in this context (see {\it e.g.}\hspace{1.1 mm}references~\cite{First1}). However, to our knowledge, this connection has not yet been exploited to compute vacuum refractive indexes \emph{directly} from the forward light-light scattering amplitude -- at least in the case where the polarization is produced by a single, `static' field as in the case of the Kerr and the Cotton-Mouton birefringences.\\ 

Our goal is to obtain the refraction indexes in various polarization configurations by using not the Lagrangian of Eq.\hspace{1.1mm}(1), but the forward light-light scattering amplitude. This approach brings out the similarity between birefringence in vacuum and birefringence in material media.\\

In the following section, we will recall the classic relation between refractive indexes and forward scattering amplitudes. Next, we will recall the expression of these amplitudes in the case of light-light scattering. These results will first be used to compute the phase velocity of two (intense) head-on polarized light waves. Then, we will show how this formalism can be extended to `static' fields like those which produce the Kerr and the Cotton-Mouton effects. Finally, we will treat a few other field configurations that involve both an electric field and a  magnetic one.

\section{Relation between the refractive index and the forward scattering amplitude}

Let us consider a plane wave $\psi$ of unit amplitude whose quanta of energy $\omega$ are supposed to have zero mass and no spin for simplicity. We suppose that $\psi$  propagates in the positive direction of a $z$ axis and we assume that the $z<0$ region is the vacuum, while the $z>0$ region is a medium (here supposed at rest) that contains scattering centers randomly distributed with an average number density $N$. We further assume that $N$ is large enough so that the average distance between any center and its closest neighbours is much smaller than the wavelength $\lambda = 2 \pi/k$ of the incident wave. Then, in the $z<0$ region, $\psi = e^{i(\omega t - kz)}$, while in the $z>0$ region, $\psi = e^{i(\omega t - nkz)}$, where $n$ is the refractive index of the medium. Assuming that the \emph{elastic} scattering of $\psi$ by any of the scattering centers gives rise to a wave of the form $ f(\theta) e^{i(\omega t -kr)}/r$, one may establish~\cite{Hamilton}) the relation

\begin{equation} 
n = 1 + \frac{2 \pi }{\omega^2} N f(0).
\end{equation}

In the following, we will be concerned with the photon-photon scattering amplitude. Quantum electrodynamics allows one to obtain the amplitude $f(\theta)$ in the form of a series of terms of order $\alpha^n$, the first non-vanishing one being of order $\alpha^2$. As long as the energy of the colliding photons is too low for $e^+$- $e^-$ pair production, this first term is real for $\theta =0$. Thus we will be dealing with real refractive indexes. The fact that the \emph{imaginary} part of the forward scattering amplitude, $\Im[f(0)]$, vanishes to order $\alpha^2$ follows from the optical theorem, which states that  $\Im[f(0)]$ is proportional to the total cross section, $\sigma_{elastic} + \sigma_{inelastic}$. At energies below the $e^+$- $e^-$ pair production threshold, $\sigma_{inelastic}=0$ since there is no open inelastic channel, and  $\sigma_{elastic}$ is of order $\alpha^4$ since the elastic (differential) cross section is proportional to $|f(\theta)|^2$.

\section{Photon-photon forward scattering amplitudes}

In the case of photon-photon scattering, because these quanta carry a spin 1 and are observed in two different helicity states, five independent amplitudes are necessary to cover all scattering configurations. The corresponding matrix elements can be read in an article by Karplus and Neuman~\cite{KandN}.\\

Since the experimental investigation of  birefringence effects is done with linearly polarized light waves, it is more convenient to use such waves to describe the polarization state of the incoming and the outgoing photons. The matrix elements expressed in such a basis can be readily calculated from those given in a helicity basis. One may also obtain them directly from  Akhieser and Berestetskii~\cite{AandB} when the center of mass energy is low compared to the mass of an $e^+$- $e^-$ pair.\\

In fact we only need the forward \emph{elastic}  amplitudes with no spin-flip. There are two of them: $f_{\parallel}(0)$ and $f_{\perp}(0)$. They correspond to incoming and outgoing waves whose electric fields oscillate in the same plane ($f_{\parallel}(0)$) or in orthogonal planes ($f_{\perp}(0))$. Some numerical and kinematical factors must be introduced to translate the matrix elements that enter the cross section calculations into forward scattering amplitudes. Referring specifically to the matrix elements given by Akhieser and Berestetskii for the $1 + 2 \rightarrow 3 + 4$ photon-photon scattering process, one finds that, in the center of mass frame,

\begin{equation}
f_{\parallel}(0, \omega) = \frac{\alpha^2}{8\pi}\omega^3 M^{1\ 2\ 3\ 4}_{\perp\,\perp\,\perp\,\perp}(0, \omega) = \frac{\alpha^2}{4 \pi} \frac{\omega^3}{m_e^4} \frac{32}{45}, 
\end{equation}

\begin{equation}
f_{\perp}(0, \omega) = \frac{\alpha^2}{8\pi}\omega^3 M^{1\ 2\ 3\ 4}_{\perp\,\,\parallel\,\perp\,\,\parallel} (0, \omega) = \frac{\alpha^2}{4 \pi} \frac{\omega^3}{m_e^4} \frac{56}{45},
\end{equation}

\noindent
where $\omega$ is the common energy of the colliding photons, and $M^{1\ 2\ 3\ 4}_{\perp\,\perp\,\perp\,\perp}(\theta, \omega)$ (respectively  $M^{1\ 2\ 3\ 4}_{\perp\,\,\parallel\,\perp\,\,\parallel} (\theta, \omega)$) is the matrix element that corresponds to the case where all four waves are linearly polarized perpendicularly to the scattering plane (resp.\hspace{1.1mm}the case where waves 1 and 3 are linearly polarized perpendicularly to the scattering plane and waves 2 and 4 are linearly polarized in the scattering plane).

\section{Vacuum polarization produced by an intense laser beam}

The forward scattering amplitudes given by Eqs.\hspace{1.1mm}(3) and (4) are all we need to apply Eq.\hspace{1.1mm}(2) to various polarization schemes. We first take up the case where the polarization is produced by an intense laser beam as it was first analysed by Aleksandrov  \emph{et al.}~\cite{Aleksandrov}.\\

Let two intense laser beams propagate in opposite directions. We assimilate them to two plane waves and we suppose that they are linearly polarized in the $\hat{\rm{\mathbf{e}}}_x$ direction. For definiteness, we suppose that wave (1) propagates in the positive $z$ direction (like $\psi$ above) while wave (2) propagates in the opposite one. The electric fields of these two waves are then ${\mathrm{\textbf{E}}}_1 = E_1 e^{i(\omega t -kz +\varphi_1)} \hat{\rm{\mathbf{e}}}_x$ and ${\rm{\mathbf{E}}}_2 = E_2 e^{i(\omega t +kz +\varphi_2)} \hat{\rm{\mathbf{e}}}_x$, assuming that the observations are made in the center of mass frame. $\varphi_1$ and $\varphi_2$ are arbitrary phases.\\

Each wave is scattered by the photons of the opposite one and thus has a phase velocity that differs very slightly from  $c$. Let us apply Eq.\hspace{1.1mm}(2) to the propagation of wave (1) (just the same analysis could be applied to wave (2)), ignoring high order corrections due to the fact that wave (2) is affected -- very little -- by the presence of wave (1).  For $N$ one must take two times the space-averaged number density of photons per unit volume in wave (2) :

\begin{equation}
N = 2 \times(E_2)^2/2 \omega.
\end{equation}

The factor 2 introduced in this expression of $N$ is best understood in the language of particle physics. Indeed, we  may assimilate the set of scattering centers to a `target' through which wave (2) is propagating. Then the relevant parameter is the target thickness $\mathcal{T}$ crossed per unit time by the latter wave. If the scattering centers were at rest, $\mathcal{T}$ would be equal to $cN$, but since the target photons associated with wave (2) move with a velocity nearly equal to $c$ in the $z<0$ direction, the value of $\mathcal{T}$ is doubled.\\

Inserting (3) and (5) into Eq.\hspace{1.1mm}(2), we obtain

\begin{equation}
n_\parallel = 1 +\frac{16}{45}\alpha^2 \frac{(E_2)^2}{m_e^4}.
\end{equation}

If waves (1) and (2) are  polarized in orthogonal planes, the numerical coefficient in Eq.\hspace{1.1mm}(6) must be replaced by $\frac{28}{45}$ (cf. Eq.\hspace{1.1mm}(4)). Thus

\begin{equation}
n_\perp = 1 +\frac{28}{45}\alpha^2\frac{(E_2)^2}{m_e^4}.
\end{equation}

\noindent 
One notices that while the light-light forward scattering amplitude has a strong dependence upon the energy $\omega$, the three powers of $\omega$ that enter the expression (2) of the refractive index cancel out.\\ 

These results are in agreement with those of  Aleksandrov  \emph{et al.}~\cite{Aleksandrov}.\\

Up to now, we have assumed that we were working in the center of mass of the colliding photons but Eqs.\hspace{1.1mm}(6) and (7) remain valid when waves (1) and (2) have different frequencies. Indeed, if the observations are made in a frame that moves in the $z$ direction with a velocity $\beta$ with respect to the C of M frame\footnote{Lorentz transformations are extensively exploited in Ref.\hspace{1.1mm}[6]}, the electric field of wave (2) becomes $(E'_2)^2  = [(1+\beta)/(1-\beta)](E_2)^2$, while the new phase velocity of wave (1) satisfies $v' = (v-\beta)/(1 + v \beta).$ The latter relation implies $(n'-1) = [(1+\beta)/(1-\beta)](n-1)$ since the product $(n-1)(n'-1)$ may be neglected compared to $n-1$ or $n'-1$.

\section{Kerr and Cotton-Mouton birefringences in vacuum}

One may wonder how this light-light scattering approach can be extended to the Kerr or the Cotton-Mouton effects since they involve static fields, ${\rm{\mathbf{E}}}_0$ or ${\rm{\mathbf{B}}}_0$, and only one of them at a time. The reason why it can be done is that the wavelength of the polarizing wave  -- \emph{i.e.} wave (2) in the preceding section -- does not enter  Eqs.\hspace{1.1mm}(6) or (7). Thus one may assume that the static field is in fact produced by an electromagnetic wave propagating in the \emph{negative} $z$ direction and whose  wavelength is long enough to create an electric field (or a magnetic one) as uniform as one wishes within the experimental apparatus. Of course, a single wave would create both an electric and a magnetic field, but either one can be cancelled by assuming the presence of a second wave of the same amplitude but propagating in the \emph{positive} $z$ direction.\\

Let us apply this procedure to the Kerr effect. We assume that it is  produced  in the vicinity of the origin  of our reference frame by an electric field  ${\rm{\mathbf{E}}_0} = E_0 \hat{\rm{\textbf{e}}}_x $. We may suppose such a field results (near the origin, \emph{i.e.} for $z \approx 0$) from the presence of the  following two waves: wave (2) whose electric field is  ${\rm{\mathbf{E}}}_2 = (E_0/2) e^{i(\Omega t +Kz +\phi)} \hat{\rm{\textbf{e}}}_x$, and wave (2') whose electric field is ${\rm{\mathbf{E}}}_{2'} = (E_0/2) e^{i(\Omega t -Kz +\phi)} \hat{\rm{\textbf{e}}}_x$, at the time $t$ such that  $\Omega t+ \phi \approx 0$ (one could relax this condition by making a different choice of the waves' common amplitude). The wave number $K$ is chosen small enough -- as well as the wave frequency $\Omega = Kc$ --  so that the field uniformity is adequate within the experimental apparatus. The combination of these two waves provides the desired electric field while the total magnetic field is vanishing.\\

Clearly, the photons of wave (2') do not scatter wave (1) since they move in the same direction. Therefore birefringence effects will only come from wave (2) and one may apply the results of the previous section with the following changes in Eqs.\hspace{1.1mm}(6) and (7): The amplitude $E_0/2$ must be substituted to the amplitude $E_2$, and the numerical factors must be multiplied by 2. The reason for this factor 2 is the following. Photons are statistically distributed in space proportionnally to the square of the local amplitude of the electromagnetic field; since we chose $\Omega t +\phi \approx 0$, at that particular time the number density of photons  near the origin is twice its space-averaged value. Thus, for the Kerr effect,

\begin{equation}
n_\parallel = 1 + \frac{8}{45}\alpha^2 \frac{(E_0)^2}{m_e^4}\:\:\:{\rm{and}}\:\:\: n_\perp = 1 + \frac{14}{45}\alpha^2 \frac{(E_0)^2}{m_e^4}.
\end{equation}

The same analysis can be applied to the Cotton-Mouton effect produced by a uniform magnetic field  ${\rm{\mathbf{B}}_0}$ oriented perpendicularly to the wave vector $k\hat{\rm{\mathbf{e}}}_z$. Consequently, in order to obtain the refractive indexes of the vacuum thus polarized, one has merely to substitute $B_0$ to $E_0$ and to perform the interchange $ \frac{8}{45} \leftrightarrow \frac{14}{45}$ in Eqs.\hspace{1.1mm}(8). This interchange reflects the fact that, to produce ${\rm{\mathbf{B}}_0}$, one has to combine two waves whose electric fields are at right angle with the direction of ${\rm{\mathbf{B}}_0}$.\\

These results reproduce \emph{e.g.} those of  Rikken and Rizzo~\cite{Rikken1}.

\section{More complex polarization schemes}

We now consider the case treated by Rikken and  Rizzo where the vacuum is polarized by a combination of `static' electric and magnetic fields. We first assume that these fields are mutually orthogonal:  ${\rm{\mathbf{E}}}_0 = E_0 \hat{\rm{\textbf{e}}}_y$, ${\rm{\mathbf{B}}}_0 = B_0 \hat{\rm{\textbf{e}}}_x $. Note that, in accordance with~\cite{Rikken1}, our polarizing fields are chosen so that the \{$\rm{\mathbf{E}}_0$,  $\rm{\mathbf{B}}_0$, $\hat{\mathbf{e}}_z$\} frame is \emph{left}-handed when $E_0$ and $B_0$ have the same sign.\\

Let us assume for a moment that $E_0 = - B_0$. Then the two `static' fields can be produced by a single wave -- with a sufficiently long wavelength -- that travels in the \emph{positive} $z$ direction so that its photons do not collide with those of wave (1). We can immediately conclude that such a field configuration does not generate any birefringence (to order $\alpha^2$). This striking result also follows from Rikken and Rizzo's expressions for $n_\parallel$ and $n_\perp$ but, to our knowledge, it has not yet been pointed out nor interpreted so simply.\\

We now turn to the general case where $E_0$ and $B_0$ are not necessarily equal in magnitude. We assume for definiteness that both amplitudes are positive and that  $E_0 > B_0$ (other cases can be treated in just the same way) and let $\Delta$ be the difference $E_0$ - $B_0$. \\

Such a field configuration can be produced by the superposition of three waves\footnote{This multiwave approach should not be confused with the so-called `Four-wave mixing' scheme utilized in non-linear optics where three electromagnetic waves interact to generate a fourth one whose characteristics follow from energy-momentum conservation.}, two of them  (waves (2) and (3)) that propagate in the $z<0$ direction and one (wave (3')) that propagates in the $z>0$ direction. Their electric fields are, respectively, ${\rm{\mathbf{E}}}_2 = B_0 e^{i(\Omega t +Kz +\phi)} \hat{\rm{\textbf{e}}}_x$,  ${\rm{\mathbf{E}}}_{3} = (\Delta/2) e^{i(\Omega t +Kz +\phi)} \hat{\textbf{e}}_x$ and  ${\rm{\mathbf{E}}}_{3'} = (\Delta/2) e^{i(\Omega t -Kz +\phi)} \hat{\rm{\textbf{e}}}_x$.\\

As done above, we  suppose that the birefringence effect is observed near the origin of our frame of reference ($z\approx 0$) at the time $t$ such that $\Omega t+\phi \approx 0$. Wave (3') travels in the same direction as wave (1) and thus does not generate any  birefringence, while waves  (2) and (3) combine coherently in a single wave of amplitude $B_0 + \Delta/2 = (E_0 + B_0)/2$. According to Eqs.\hspace{1.1mm}(6) and (7), the latter wave generates a birefringence effect with refractive indexes

\begin{equation}
n_\parallel = 1 + 2\times\frac{4}{45}\alpha^2 \frac{(E_0+B_0)^2}{m_e^4}\:\:\: {\rm{and}}\:\:\: n_\perp = 1 + 2\times\frac{7}{45}\alpha^2 \frac{(E_0+B_0)^2}{m_e^4}.
\end{equation}

\noindent 
Note that factors $2$  have been inserted in Eqs.\hspace{1.1mm}(9) to take into account the local value of the photon number density.\\

Again these results agree with those of Rikken and Rizzo. For $|E_0| \neq  |B_0|$ they can also be derived from Eqs.\hspace{1.1mm}(8) by performing a Lorentz transformation along the $z$ axis.\\

Finally, let us consider the case where the `static' electric and magnetic fields that polarize the vacuum are parallel. The most general configuration where $E_0 \neq B_0$ can be treated along the same lines as above. Here we will limit ourselves to the case where  $E_0 = B_0 $ and assume that these fields are both in the $\hat{\textbf{e}}_x$ direction. This field configuration may be regarded as resulting from the superposition of four electromagnetic waves, all of amplitude $E_0/2$:  ${\rm{\mathbf{E}}}_2 = (E_0/2) e^{i(\Omega t +Kz+\phi)} \hat{\rm{\textbf{e}}}_x$, ${\rm{\mathbf{E}}}_{2'} = (E_0/2) e^{i(\Omega t -Kz + \phi)} \hat{\rm{\textbf{e}}}_x$, ${\rm{\mathbf{E}}}_3 = (E_0/2) e^{i(\Omega t +Kz+\phi)} \hat{\rm{\textbf{e}}}_y$ and ${\rm{\mathbf{E}}}_{3'} = (E_0/2) e^{i(\Omega t -Kz+\phi)} \hat{\rm{\textbf{e}}}_y$. Two of these waves  propagate in the $z>0$ direction and therefore do not interact with wave (1). The other two propagate in the $z<0$  direction and are polarized along $\hat{\textbf{e}}_x$ for one and $\hat{\textbf{e}}_y$ for the other. The latter two waves add coherently to produce a single wave of amplitude $E_0/\sqrt{2}$ oriented at $\pi/4$ with respect to the $\hat{\textbf{e}}_x$, $\hat{\textbf{e}}_y$ axes. We conclude that such a field configuration leads to birefringence axes oriented at $\pm \pi/4$ with respect of the direction common to the polarizing fields.

\section{Conclusion}

We have shown how vacuum birefringence can be dealt with in the framework of coherent scattering as is done for refraction effects in dense material media. When the vacuum is polarized by an intense light wave, we assign to the wave's photons the role of the `dense medium'. We have further shown how this procedure can be extended to the case where the vacuum is polarized by a static electromagnetic field. This is achieved by supposing that the static field is actually produced by a superposition of waves that propagate in opposite directions. With this approach, birefringence indexes previously calculated by other authors are obtained in a straightforward way for various polarization schemes. It is hoped that a new insight into the process that generates birefringence in vacuum is thus provided.

\end{document}